\documentclass[conference]{IEEEtran}
\IEEEoverridecommandlockouts
\usepackage{cite}
\usepackage{amsmath,amssymb,amsfonts}
\usepackage{algorithm, algorithmic}
\usepackage{graphicx}
\usepackage{textcomp}
\usepackage{xcolor}
\usepackage{comment}
\def\BibTeX{{\rm B\kern-.05em{\sc i\kern-.025em b}\kern-.08em
    T\kern-.1667em\lower.7ex\hbox{E}\kern-.125emX}}
\begin{document}

\title{Semantic-Native Communication: A Simplicial Complex Perspective}

\author{
\IEEEauthorblockN{Qiyang Zhao\IEEEauthorrefmark{1}, Mehdi Bennis\IEEEauthorrefmark{2}, Mérouane Debbah\IEEEauthorrefmark{1}, Daniel Benevides da Costa\IEEEauthorrefmark{1}}
\IEEEauthorblockA{\IEEEauthorrefmark{1}Technology Innovation Institute, 9639 Masdar City, Abu Dhabi, UAE}
\IEEEauthorblockA{\IEEEauthorrefmark{2}Centre for Wireless Communications, University of Oulu, Oulu 90014, Finland}
Email: qiyang.zhao@tii.ae, mehdi.bennis@oulu.fi, merouane.debbah@tii.ae, daniel.costa@tii.ae
}
\maketitle

\begin{abstract}
Semantic communication enables intelligent agents to extract meaning (or semantics) of information via interaction, to carry out collaborative tasks. In this paper, we study semantic communication from a topological space perspective, in which higher-order data semantics live in a simplicial complex. Specifically, a transmitter first maps its data into a $k$-order simplicial complex and then learns its high-order correlations. The simplicial structure and corresponding features are encoded into semantic embeddings in latent space for transmission. Subsequently, the receiver decodes the structure and infers the missing or distorted data. The transmitter and receiver collaboratively train a simplicial convolutional autoencoder to accomplish the semantic communication task. Experiments are carried out on a real dataset of Semantic Scholar Open Research Corpus, where one part of the semantic embedding is missing or distorted during communication. Numerical results show that the simplicial convolutional autoencoder enabled semantic communication effectively rebuilds the simplicial features and infer the missing data with $95\%$ accuracy, while achieving stable performance under channel noise. In contrast, the conventional autoencoder enabled communication fails to infer any missing data. Moreover, our approach is shown to effectively infer the distorted data without prior simplicial structure knowledge at the receiver, by learning extracted semantic information during communications. Leveraging the topological nature of information, the proposed method is also shown to be more reliable and efficient compared to several baselines, notably at low signal-to-noise (SNR) levels.
\end{abstract}


\section{Introduction}
Semantic and goal-oriented communications are seen as a  promising technology going beyond transferring and reconstructing information bits (Shannon's level-A). In contrast, extracting and understanding the meaning of information (Shannon's level-B) is instrumental in effectively solving a task or goal (Shannon's level-C) \cite{Seo2021}. In doing so, higher communication efficiency and reliability can be achieved by leveraging the agents' ability to reason and infer missing and distorted data, without the need to accurately communicate and reconstruct the original data. 

Semantic communication (SC) brings significant challenges to the communications system design, from both theoretical and practical standpoints. To tackle these issues, significant research works have evolved rapidly in the recent years. From a Machine Learning (ML) viewpoint, current works are focused on applying off-the-shelf techniques including convolutional neural network (CNN), long short-term memory (LSTM), transformers, autoencoders and others. All of which are based on extracting latent features from a given input (text, images, etc) and communicate them to a receiver. For instance, LSTM and transformer have shown success in extracting semantic information from text messages \cite{Xie2021}, borrowing the bilingual evaluation understudy (BLEU) score as a semantic metric, compared to conventional source (Huffman) channel (Turbo) coding, especially at low signal-to-noise (SNR) values \cite{Seo2021}. In the context of semantic channel coding, an adaptive universal transformer was proposed in \cite{Zhou2021} and \cite{Lu2022} by using an autoencoder to encode text sentences into high dimensional latent representations, and channel state information (CSI) was used to adjust attention weights according to the SNR regime. The work in \cite{Jiang_2022} incorporated hybrid automatic repeat request (HARQ) to transmit incremental bits according to reliability. Another application of non-linear transformer on images is proposed in \cite{Dai2021}, by mapping source images in latent representation and learning the entropy model of latent distribution. Finally, \cite{Weng2022} studied the problem of speech recognition and synthesis with joint CNN and recursive neural network (RNN) modules, to recover the text and speech at the receiver. While interesting, these works focus on learning latent representations directly from raw data, to compress data at the transmitter and reconstruct it at the receiver. 

A different line of work casts the problem of semantic communication as a belief transport problem among agents that reason over one another, to communicate only the minimum amount of semantic information. Building on this framework, the proposed idea in our work is based on first mapping a set of observations  (raw data) onto a topological (concept) space before  learning and infering high-order information using ML techniques. Due to the non-structured nature of data, we leverage recent advances in topological/geometric deep neural networks \cite{Bronstein_2017}, \cite{Feng2018} that operate on non-Euclidean spaces. In particular,  simplicial and cell complexes \cite{Ebli2020}, \cite{Hajij2020} can learn higher-order interactions beyond pairwise elements in a graph. 

The key contribution of this work can be summarized as follows. Firstly, we introduce a topological perspective for SC rooted in mapping unstructured raw data onto  simplicial complexes. Secondly, in a transmitter-receiver setting, we use a simplicial convolutional network (SCN) at the transmitter to extract hidden high-order patterns and correlations between simplices at different levels of abstractions. Then,  both  simplicial complex  topology and features are embedded into a latent vector for transmission to a receiver. Thirdly, we propose a simplicial autoencoder (SAE) to reconstruct the simplicial topology at the receiver, and infer the missing or distorted data features on the simplicial complex. The transmitter and receiver pair trains the SAE by iteratively minimizing a reconstruction error of the simplicial complex. We validate the communication effectiveness of the proposed approach with respect to reliability and convergence under different SNR values and levels of abstractions (via the embedding length). The proposed method not only compresses data but also infers higher-order correlations at both transmitter and receiver sides. To the best of our knowledge, this is the first work studying semantic communications from a high-order simplicial complex perspective. 

The rest of this paper is structured as follows. Section \ref{sec:semantic_model} introduces the SC system model on simplicial complexes. Section \ref{sec:simplicial_model} describes the simplicial convolutional autoencoder for semantic coding and communication. Section \ref{sec:perforamnce} provides numerical results of the proposed approaches on a real dataset along with insightful discussions. Finally, the work is concluded with several future directions discussed in Section \ref{sec:conclusion}. The notations used in the article are summarized in Table \ref{tab:notations}.
\begin{table}
	\centering
	\caption{Notations in Simplicial Autoencoder Formulation}
	\label{tab:notations}
	\begin{tabular}{|c|l|}
		\hline
		\textbf{Notation} & \textbf{Description} \\ \hline
        $S^k$ & $k$-order simplicial complex with simplices $s \in S_k$\\ \hline
        $\mathcal{L}$ & Hodge Laplacian matrix of simplicial complex \\ \hline
        $X$ & Feature on simplicial complex, $x \in X(S_k)$ \\ \hline
        $C$ & Cochains of simplicial complex, $c \in C(S_k)$ \\ \hline
        $\pi$ & Linear coboundary operator \\ \hline
        $B$ & Sparse matrix of coboundaries \\ \hline
        $\delta$ & Nonlinear activation function with bias $b$ \\ \hline
        $V$ & Semantic embedding of simplicial complex \\ \hline
        $f$ & Hidden features of simplicial cochains \\ \hline
        $\lambda$ & Eigenvalues of simplicial Laplacians \\ \hline
        $\mathcal{F}$ & Fourier transform of simplicial Laplacians \\ \hline
        $\hat{S}$ & Received simplices, $\{s \in S\}$ \\ \hline
        $\hat{S}^c$ & Missing or distorted simplices, $\{s \in S: s \notin \hat{S}\}$ \\ \hline
        $S'$ & Inferred simplices from simplicial autoencoder \\ \hline
        $E_\psi$ & Simplicial complex encoder at transmitter \\ \hline
        $D_\phi$ & Simplicial feature decoder at receiver \\ \hline
        $D_\varphi$ & Simplicial structure decoder at receiver \\ \hline
        $J(\cdot)$ & Objective funciton \\ \hline
        $\alpha$ & Learning rate\\ \hline
        $t$ & Communication round\\ \hline
        $|\cdot|$ & Number of elements in a set (i.e. complexes, features) \\ \hline
        $*$ & Simplicial convolution \\ \hline
	\end{tabular}
\end{table}

\section{Simplicial Complex Model} \label{sec:semantic_model}

\begin{figure}
	\centering
			\setlength{\abovecaptionskip}{0pt}
	\includegraphics[width=\linewidth]{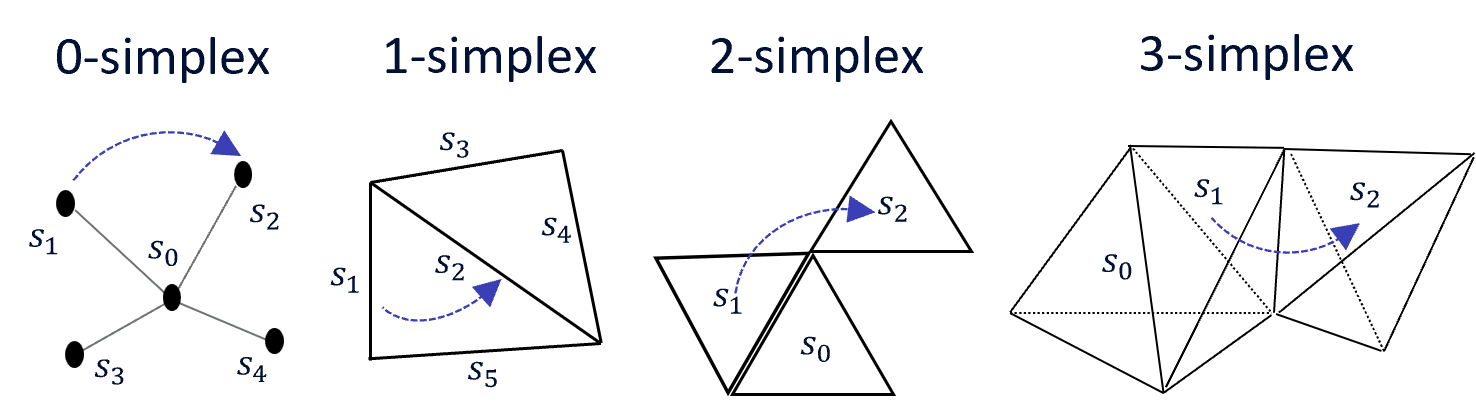}
	\caption{Geometric interpretation of a $k$-order simplicial complex (arrow indicates a $k-$Laplacian).}
	\label{fig:simplicial_complex}
	\vspace{-12pt}
\end{figure}

The semantic communication problem under consideration aims to extract and transmit semantic representations of data on simplicial complexes over a wireless channel. The goal is to train a receiver to reconstruct semantically similar simplicial complexes, and infer missing and distorted data during communication. For validation, we consider a co-authorship dataset \cite{Ebli2020}, and learn semantic representations of this dataset on $k$-order simplicial complexes. A $k$-simplex is a topological/geometric object with $(k+1)$ vertices, containing $(k+1)$ faces of dimension $(k-1)$, denoted as $\{v_0,...,\hat{v}_i,...,v_k\}$ \cite{Ebli2020}. If a simplex $s$ is a face of $\tau$, then $\tau$ is called a coface of $s$. Such combinatorial definition induces a geometric interpretation, in that $0$-simplices are vertices, $1$-simplices are edges, $2$-simplices are triangles, $3$-simplices are tetrahedrons, and so on (see Fig. \ref{fig:simplicial_complex}). A simplicial complex $S_k$ is thus a collection of simplices $s$ with a dimension of $k$, that is closed under inclusion of all faces. Effectively, given out dataset we map a paper with $k$ authors to a $(k-1)$-simplex, in which subsimplices represent papers from a subset of authors of this paper. With the paper's features (i.e. citations, references, subjects) modeled on a simplex, the higher-order semantic information of the corpus can be captured by simplicial complex. 

We define $C_k(S)$ as the $k$-cochains of simplicial complex $S$ in $\mathbb{R}$-vector space. The linear coboundary operator for each order $k$ can be defined as:
\begin{equation}
	\label{eq:coboundary}
	\begin{gathered}
    \pi_k:C_k(S)\rightarrow C_{k-1}(S) \\
    \pi_k(f)([v_0,...,v_k])=\sum_{i=0}^{k}(-1)^k f([v_0,...,\hat{v}_i,...,v_k]),
    \end{gathered}
\end{equation}
\noindent where the support of hidden feature $f$ of cochains $C$ is constructed from cofaces of $k$-simplices. 

For a finite  simplicial complex, the degree $k$ of the Hodge  Laplacian of a simplicial complex $S_k$ is defined as a linear map of its adjoint of the coboundary $\pi^{k*}$ \cite{Belfiore2021}:
\begin{equation}
	\label{eq:laplacian}
	\begin{gathered}
	\mathcal{L}_k:C^k(S)\rightarrow C^k(S) \\
	\mathcal{L}_k = \pi^{k*} \circ \pi^k + \pi^{(k - 1)} \circ \pi^{(k - 1)*},
    \end{gathered}
\end{equation}

The coboundary can be represented as a sparse matrix $B_i$, and the Laplacian can be computed as
\begin{equation}
	\label{eq:coboundary_sparse}
	\begin{gathered}
	\mathcal{L}_k = B_k^T B_k + B_{k - 1} B_{k - 1} ^ T,
    \end{gathered}
\end{equation}

\noindent where $\mathcal{L}_0$ and $B_0$ are the classic graph Laplacian and incidence matrix, respectively. The kernel of the $k$-Laplacian is isomorphic to the $k$-(co)homology of its associated simplicial complex \cite{HORAK2013303}. 

The $k$-Laplacian matrix $\mathcal{L}_k$ is constructed by papers with $k$ coauthors. The features of interest (also referred to as cochains) of the $(k-1)$-simplex $S_{k-1}$ encode the total number of citations of papers with $k$ authors. In short, the coauthorship dataset is projected onto an undirected unweighted featured simplicial complex $S$, containing $|S|$ simplices with $|B|$ coboundaries, and each simplex has a $|X|$ dimensional feature vector. Thus, the simplicial complex is characterized by the Laplacian and feature matrix $S(\mathcal{L}, X)$.

\section{Simplical Convolutional Autoencoder enabled Semantic Communication} \label{sec:simplicial_model}

Studying the problem of semantic communication through the lens of simplicial complexes is done as follows: the transmitter extracts high-order semantic information in $S_k$ using an SCN, and encodes both the simplicial structure $\mathcal{L}$ and feature $X$ into a latent vector $V$ for communication purposes. The receiver's task is to first decode the structure, and then infer the missing or distorted features using an SCN. In each communication round, the SAE is trained to approximate the original simplicial complex.

\subsection{Simplicial Convolution}
The simplicial convolution layer used in this work extends the notion of convolution over a grid (as in CNN) towards non-Euclidean data as $\delta \circ (f*\varphi_w)$, where $\delta$ is a nonlinear activation function with bias, $\varphi_w\in R^{|S_k|}$ represents a low-degree polynomial filter in terms of the eigenvalues of the ${k}$-Laplacians $\lambda_k$, parameterized by trainable weights $W$ \cite{Ebli2020}:
\begin{equation}
	\label{eq:filter}
	\begin{gathered}
	\varphi_w = \sum_{i = 0}^{|S_k|}W_i\Lambda^i = \sum_{i = 0}^{|S_k|}W_i(\lambda_1^{i}, \lambda_2^{i}, ..., \lambda_{|S_k|}^{i}),
    \end{gathered}
\end{equation}

The simplicial convolution  is derived from the discrete Fourier transform of the Laplacians $\mathcal{F}:C_k(S)\rightarrow \mathbb{R}^{|S_k|}$, where $\mathcal{F}_k(c)$ is a set of inner product of cochains and its eigenvalues:
\begin{equation}
	\label{eq:fourier}
	\begin{gathered}
	\mathcal{F}_k(c) = \big(\langle c, \lambda_i \rangle _k \big),
    \end{gathered}
\end{equation}

The cochain convolution transformed from spectral domain is given as:
\begin{equation}
	\label{eq:convolution}
	\begin{gathered}
	c*c' = \mathcal{F}_k^{-1}(\mathcal{F}_k(c)\mathcal{F}_k(c')), \quad \forall c, c' \in C^k(S),
    \end{gathered}
\end{equation}

With this framework, we define a simplicial convolutional layer with $k-$cochain $c$ and weights $W$ as:
\begin{equation}
	\label{eq:convolution_algebra}
	\begin{gathered}
	c*c'= \delta \circ \big(\mathcal{F}_k^{-1}(\varphi_w)*c) = \delta \circ \big(\sum_{i = 0}^{N} W_i\mathcal{L}_k^ic\big),
    \end{gathered}
\end{equation}

The filter $W_i$ is restricted to simplices that are $N$ hops apart within the $k$-degree complex, where there is no message passing with its upper and lower adjacency. We apply a convolutional layer separately to each $k$-simplex to predict $N$ adjacent papers based on the input Laplacians. 

\subsection{Simplicial Autoencoder}

\begin{figure}
	\centering
			\setlength{\abovecaptionskip}{0pt}
	\includegraphics[width=\linewidth]{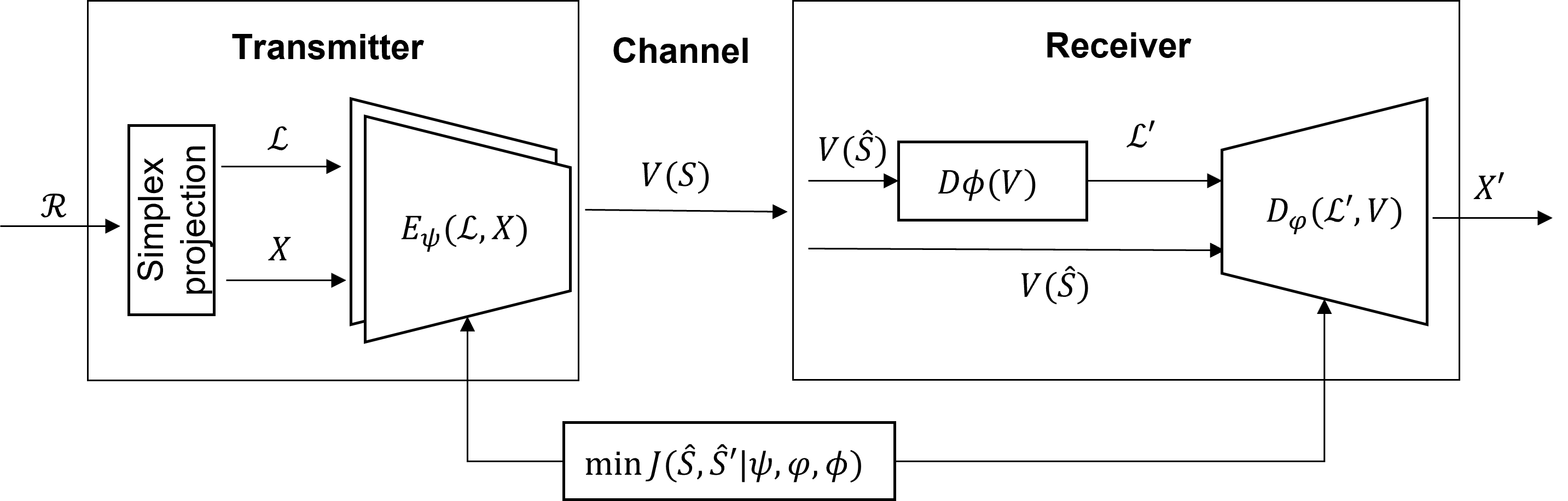}
	\caption{Simplicial Convolutional Autoencoder.}
	\label{fig:simplicial_ae}
	\vspace{-12pt}
\end{figure}

We propose an SAE for semantic communications on simplicial complexes, as illustrated in Fig. \ref{fig:simplicial_ae}. The simplicial encoder at the transmitter maps coauthorship dataset onto a simplicial complex $S(X, \mathcal{L})$, and extracts semantic embedding $V(S)$ for each simplex at $|V|$ order using a SCN. During the transmission of $V(S)$ over a noisy wireless channel with noise $\mathcal{N}$, $p\%$ simplical data are considered missing (removed) or distorted (replaced with random values), such that only a subset of simplicial embedding $V(\hat{S})$ is received. The simplicial decoder at the receiver first infers the complete simplical topology $\mathcal{L}'$, and then invokes SCN to infer the simplical feature $X'$. The transmitter and receiver then jointly train the SAE by minimizing the feature and topology error of the received simplices $\hat{S}'(\hat{X}', \hat{\mathcal{L}}')$. We then evaluate the performance of missing or distorted simplices, denoted as $\hat{S}^c = \{s \in S: s\notin \hat{S} \}$, inferred by the SAE. During training, the objective is to achieve a minimum error of both received and missing or distorted simplices. 

The encoder leverages a stacked SCN to extract the semantics of the featured complex with $|S|$ simplices and $|X|$ features each. The feature $X$ and Laplacian $\mathcal{L}$ matrix are embedded onto a latent matrix $V(S)$, with $v$ dimension per simplex. Thus the compression level of embedding is $(|S|+|X|)/|V|$, where $|S|, |X|\ll|V|$, yielding
\begin{equation}
	\label{eq:enc_scn}
	\begin{gathered}
    E_\psi:\mathcal{L}, X \rightarrow V, \\
	V_k^i = \delta \circ \big(\sum_{i = 0}^{|S_k|} W_v^k \mathcal{L}_i^k x_i + b_v^k\big), \quad \forall x_i \in X_k,
    \end{gathered}
\end{equation}

Upon receiving an incomplete simplicial embedding $V(\hat{S})$, the receiver applies a trainable weight matrix bi-linearly to every pair of simplices' latent vector $v_i, v_j$, and maps it onto the output $k$-Laplacian:
\begin{equation}
	\label{eq:dec_mlb}
	\begin{gathered}
    D_\phi: V \rightarrow \mathcal{L}, \\
	\mathcal{L}_k^{'(i, j)} = \delta \circ (v_i^T W_l^k v_j + b_l^k), \forall v_i, v_j \in V_k,
    \end{gathered}
\end{equation}

The output $L'_k$ is normalized for the decoder SCN to infer the feature of each simplex from the simplicial embedding:
\begin{equation}
	\label{eq:dec_scn}
	\begin{gathered}
    D_\varphi: V \rightarrow X, \\
	X_k^{'i} = \delta \circ \big(\sum_{i = 0}^{|S_k|} W_x^k \mathcal{L}_i^{'k} v_i + b_x^k\big), \forall v_i \in V_k,
    \end{gathered}
\end{equation}

At each iteration, when communicating a simplicial complex, the transmitter trains the complex encoder $E_\psi$, while the receiver trains its structure decoder $D_\phi$ and feature decoder $D_\varphi$, by minimizing a L2 loss (mean square error) of the feature $\hat{X}$ and Laplacian $\hat{\mathcal{L}}$ of the received simplices $\hat{S}$:
\begin{equation}
	\label{eq:loss_L}
	\begin{aligned}
    J(\psi, \phi, \varphi) = \frac{1}{|\hat{S}|} \sum_{i = 0}^{|\hat{S}|} \|D_{\phi, \varphi}(E_\psi(\hat{S}(\mathcal{L}, X)_i))  - \hat{S}(\mathcal{L}, X)_i\|^2,
    \end{aligned}
\end{equation}

To search the optimal values of $\phi, \varphi, \psi$, we use Stochastic Gradient Descend (SGD) to perform backpropagation over $\nabla J(\psi, \phi, \varphi)$. The activation fucntions $\delta$ is selected as LeakyReLU with negative slope $10^{-2}$. 

The gradients of the loss function are represented as:
\begin{equation}
	\label{eq:gradient_L}
	\begin{aligned} 
    \nabla J(\cdot) 
    & = \frac{\partial}{\partial{\varphi}} \frac{\partial{\varphi}}{\partial{\phi}} \frac{\partial{\phi}}{\partial{\psi}} \frac{1}{2} (D_{\phi, \varphi}(E_\psi(\hat{S}))  - \hat{S}_i)^2, \\
    & = (D_{\phi, \varphi} (E_\psi(\hat{S})) - \hat{S}) \frac{\partial}{\partial{\varphi}} \frac{\partial{\varphi}}{\partial{\phi}} \frac{\partial{\phi}}{\partial{\psi}} (\sum_{i = 0}^{|S|} \phi_i \varphi_i \psi_i \hat{S}_i -  \hat{S}), \\
    & = (D_{\phi, \varphi}(E_\psi(\hat{S})) - \hat{S}) \hat{S},
    \end{aligned}
\end{equation}

Optimization of $\phi, \varphi, \psi$ is performed iteratively by SGD with a learning rate $\alpha$:
\begin{equation}
	\label{eq:descend_L}
	\begin{aligned} 
    \psi^{t + 1}, \phi^{t + 1}, \varphi^{t + 1} = \psi^{t}, \phi^{t}, \varphi^{t} - \alpha \cdot \nabla J(\psi^{t}, \phi^{t}, \varphi^{t}) ,
    \end{aligned}
\end{equation}

Inference is evaluated over missing or distorted simplices. The error $e$ and accuracy $a$ can be defined as:
\begin{equation}
	\label{eq:mae_L}
	\begin{gathered} 
    e = \frac{1}{|\hat{S}^c|} \sum_{i = 0}^{|\hat{S}^c|} \|D_{\phi, \varphi}(E_{\psi}(\hat{S}^c_i))  - \hat{S}^c_i\|^2, \\
    a = 1 - \frac {\sum_{i = 0}^{|\hat{S}^c|} \|D_{\phi, \varphi}(E_{\psi}(\hat{S}^c_i))  - \hat{S}^c_i\|} {\sum_{i = 0}^{|\hat{S}^c|} {\hat{S}_i^c}},
    \end{gathered}
\end{equation}

The procedure of SC between transmitter (Tx) and receiver (Rx) on SAE is detailed in Algorithm \ref{alg:sae-sc}.
\begin{algorithm}
\caption{SAE-enabled Semantic Communications}
\label{alg:sae-sc}
\begin{algorithmic}[1]
\FOR {$t = 0, ..., T - 1$}
\STATE Tx performs random walk over coauthorship data set $\mathcal{R}$ and samples a simplicial complex $S(\mathcal{L}, X)$;
\STATE Tx encodes a simplicial complex into a semantic embedding $V$ based on SCN $E_\psi(\mathcal{L}, X)$;
\STATE Tx transmits $V(S)$ through a wireless channel, Rx receives a partial embedding  $\hat{V}(\mathcal{L}, S)$;
\STATE Rx infers full Laplacian $\mathcal{L}'$ based on MLB $D_\phi(V)$;
\STATE Rx infers full feature $X'$ based on SCN $D_\varphi(\mathcal{L}, V)$;
\STATE Rx computes loss of partially received simplicial complex $J(\hat{S}, \hat{S}'|\psi, \phi, \varphi)$ based on \eqref{eq:loss_L};
\STATE Rx computes gradient $\nabla J$ based on \eqref{eq:gradient_L}, updates $\phi^t, \varphi^t$ based on \eqref{eq:descend_L}, transmits $\nabla J(\phi^t, \varphi^t)$ to Tx;
\STATE Tx computes gradient $\nabla J(\psi^t)$ based on \eqref{eq:gradient_L}, updates $\psi^t$ based on \eqref{eq:descend_L};
\ENDFOR
\end{algorithmic} 
\end{algorithm}

Compared to {k}-simplex2vec and cell2vec proposed in \cite{Hacker2020} and \cite{Hajij2020}, our approach jointly embeds the feature and topology into the same latent vector. The reconstructed Laplacian is then used by the decoder SCN to infer the simplices, which is neither achieved in the joint embedding approach proposed in \cite{Lerique2019}, nor in other ML enabled SC works \cite{Xie2021}. 

\section{Performance Evaluation} \label{sec:perforamnce}
In this section, we compare the proposed method against the following benchmarks: 1) Lower bound: conventional linear autoencoder (LAE), which directly encodes the data feature without leveraging its inherent topology; 2) Upper bound: SCN \cite{Ebli2020} without semantic communication, in which the simplicial topology is assumed to be known at the receiver. We then evaluate the performance of SAE for different semantic embedding orders (compression levels) and SNR. 

\subsection{Scenario and Dataset}
We use a dataset from the Semantic Scholar Open Research Corpus \cite{Ammar2018}, which contains over 39 million research papers with authors and citations. To construct a co-authorship complex, we perform a random walk for $80$ papers with citations between 1 and 10, from a randomly sampled starting paper. A bipartie graph is created where vertices represent papers and edges connect papers sharing at least one author. Cochains between any two authors are created based on a sum citations of all collaborative papers. A set of $k$-simplicial complexes is created by joining $k$-cochains sharing authors. 

During transmission of the semantic embedding $V(S)$, we randomly remove or replace $p \%$ simplices on each complex with random values, considered as missing or distorted simplices, respectively. The receiver first rebuilds the simplices from correctly received embedding, then infers the missing or distorted simplices using SCN. 

Simulations are performed assuming an Additive White Gaussian Noise (AWGN) channel by transmitting the flattened latent vector. We compare the performance of different vector sizes with a baseline LAE on raw data. Training and testing are performed simultaneously in which an iteration represents the number of communication rounds for different complexes with a total of 500. We evaluate on an AWGN channel with SNR levels at 5, 10, 20 dB. The embedding order is set as 1, 5, 10 per simplex. The percentage of missing and distorted simplices is set to 10\%, 30\%, 50\%. The parameters are selected to test SC in different channel, distortion, redundancy scenarios. 

\subsection{Numerical Results and Discussions}
\begin{figure}
	\centering
			\setlength{\abovecaptionskip}{0pt}
	\includegraphics[width=\linewidth]{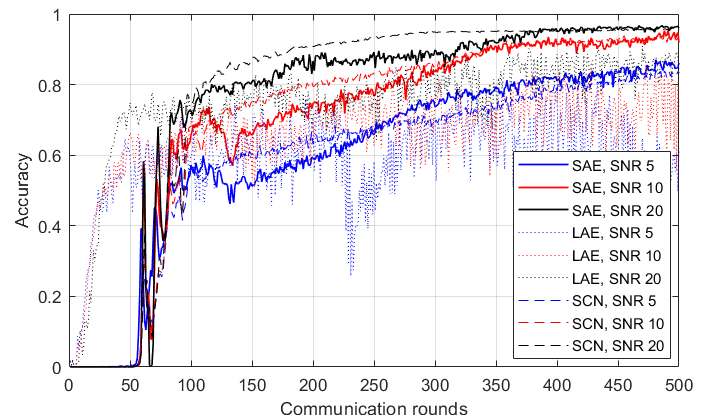}
	\caption{Simplicial feature accuracy vs benchmarks with 50\% received data}
	\label{fig:feature_accuracy_received}
	\vspace{-12pt}
\end{figure}
We first evaluate the performance of inferring simplicial features on the missing and distorted data (at 50\% of transmitted data), in which the proposed SAE approach with embedding order 10 is compared with the LAE and SCN benchmarks. The simplicial feature accuracy on received data is shown in Fig. \ref{fig:feature_accuracy_received}. It can be observed that at SNR = 20 dB, the SAE rebuilds the simplicial feature with an accuracy up to 95\%. The SCN approaches the same level with 5\% higher during training and communications. This is because SAE learns the inherent topology during iterative communications, where the SCN requires prior knowledge. Moreover, the SAE compresses data and learns to reduce the distortion, which is more communication efficient. On the other hand, the accuracy of LAE exhibits large fluctuation which is lower than SAE and SCN, even though it converges faster. This is because the LAE model lacks structure while learning and hence is affected by the channel noise, making the system unreliable.

\begin{figure}
	\centering
			\setlength{\abovecaptionskip}{0pt}
	\includegraphics[width=\linewidth]{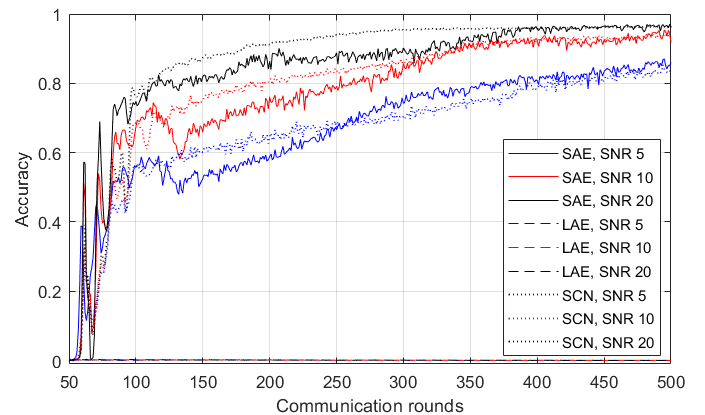}
	\caption{Simplicial feature accuracy vs benchmarks with 50\% missing data}
	\label{fig:feature_accuracy_missing}
	\vspace{-12pt}
\end{figure}

\begin{figure}
	\centering
			\setlength{\abovecaptionskip}{0pt}
	\includegraphics[width=\linewidth]{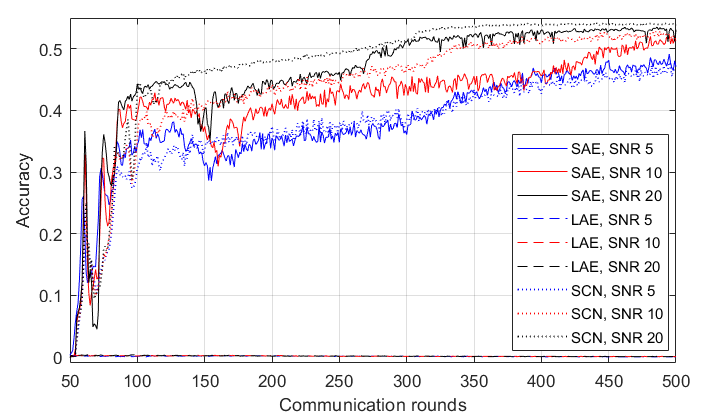}
	\caption{Simplicial feature accuracy vs benchmarks with 50\% distorted data}
	\label{fig:feature_accuracy_damage}
	\vspace{-12pt}
\end{figure}

\begin{figure}
	\centering
			\setlength{\abovecaptionskip}{0pt}
	\includegraphics[width=\linewidth]{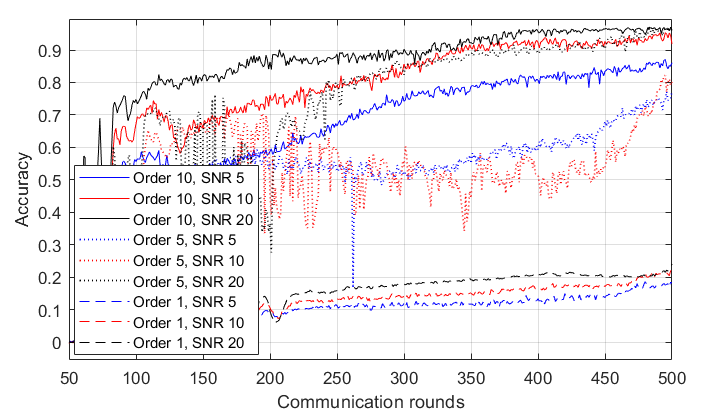}
	\caption{Simplicial feature accuracy of different semantic embedding orders with 50\% missing data}
	\label{fig:feature_accuracy_missing_order}
	\vspace{-12pt}
\end{figure}

\begin{figure}
	\centering
			\setlength{\abovecaptionskip}{0pt}
	\includegraphics[width=\linewidth]{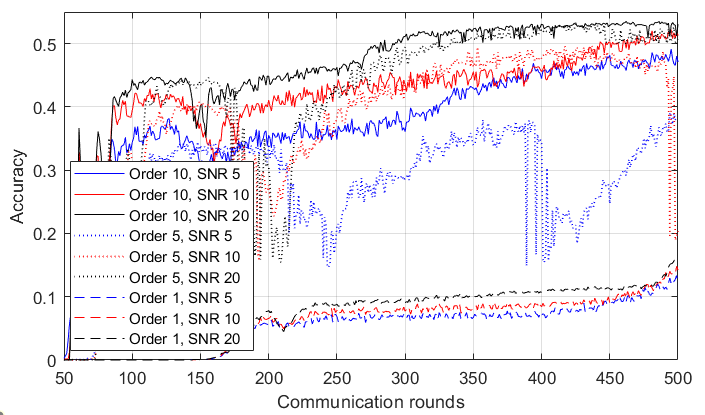}
	\caption{Simplicial feature accuracy of different semantic embedding orders with 50\% distorted data}
	\label{fig:feature_accuracy_distorted_order}
	\vspace{-12pt}
\end{figure}

The feature accuracy of missing data is shown in Fig. \ref{fig:feature_accuracy_missing}. Both SAE and SCN show similar accuracy as performed on received data. However, LAE fails to infer any missing simplices due to lack of structural information. Fig. \ref{fig:feature_accuracy_damage} shows the feature accuracy of distorted data. The maximum accuracy of SAE and SCN reaches 54\% at SNR = 20 dB. It drops by around 10\% at SNR = 5 dB where both approaches have similar accuracy. The LAE approach also fails to infer any simplices. The degradation of accuracy on distorted data compared to missing data is caused by convolution of distorted simplices with random values. According to (\ref{eq:convolution_algebra}), these simplices can further distort the adjacent correctly received simplicial embedding during SCN decoding.
We have evaluated the performance of SAE under different semantic embedding orders (denoted as $|V|$), yielding various data compression levels. Fig. \ref{fig:feature_accuracy_missing_order} shows the accuracy of the simplicial feature over 50\% missing data. With an embedding order of 5, SAE can approach the accuracy of order 10 at SNR = 20dB, though it converges slowly showing large fluctuation at a starting point below 200 rounds. However, at a SNR = 10dB the SAE with order 5 shows up to 20\% lower accuracy than SAE with order 10. Furthermore, the SAE with order 1 has low accuracy at all SNR values. The same performance metrics on distorted data is shown in Fig. \ref{fig:feature_accuracy_distorted_order}. It can be observed that the SAE with an embedding order 5 approaches the accuracy of order 10 at both SNR = 10 dB and 20 dB, performing slightly better than with missing data. However, under SNR = 5 dB it becomes unstable. We conclude from these observations that higher order semantic embeddings are instrumental in inferring missing simplices especially at low SNR.

\begin{figure}
	\centering
			\setlength{\abovecaptionskip}{0pt}
	\includegraphics[width=\linewidth]{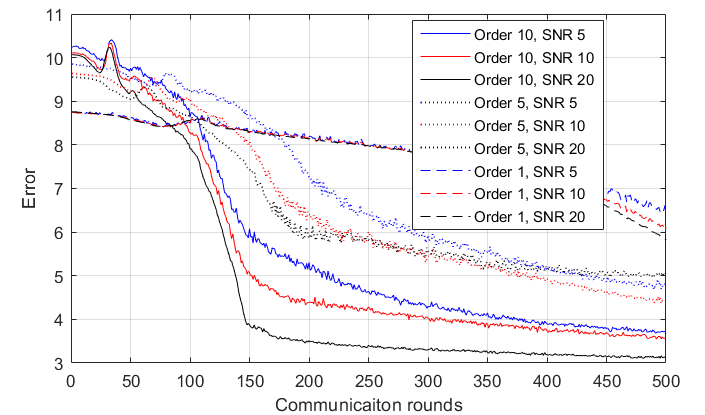}
	\caption{Simplicial topology error of different semantic embedding orders with 50\% distorted data}
	\label{fig:topology_error_distorted_order}
	\vspace{-12pt}
\end{figure}

Inference of simplicial features are based on the reconstructed topology from semantic embeddings. The absolute error of the Laplacians is shown in Fig. \ref{fig:topology_error_distorted_order}. The minimum error at 3 is achieved by embedding order 10 at SNR = 20 dB, which increases slightly under lower SNR. The embedding order 5 has higher error at all SNR levels, though its accuracy shows large difference. The embedding order 1 shows slow decrease of error, yielding low accuracy. We can see that the inference error of simplicial feature in Fig. \ref{fig:feature_accuracy_distorted_order} can be affected by the decoding error of Laplacians, and also channel variations at low SNR.

\begin{figure}
	\centering
			\setlength{\abovecaptionskip}{0pt}
	\includegraphics[width=\linewidth]{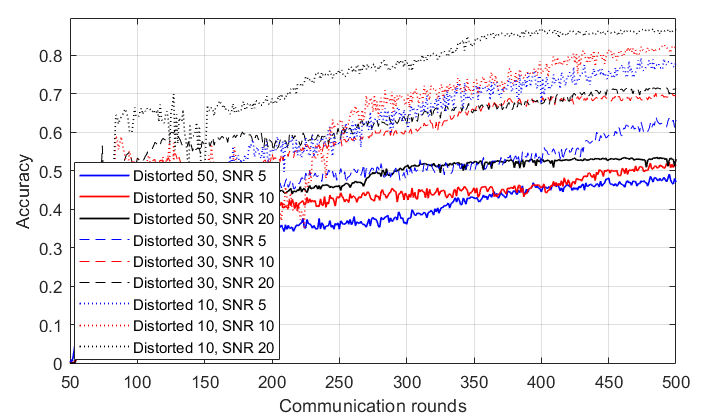}
	\caption{Simplicial feature accuracy with semantic embedding order 10 with different data distortion percentages}
	\label{fig:feature_accuracy_distorted_order10}
	\vspace{-12pt}
\end{figure}

The performance under different data distortion percentages is shown in Fig. \ref{fig:feature_accuracy_distorted_order10}, with simplicial embedding order at 10. It can be seen that at SNR = 20 dB, the SAE converges to accuracy of 50\%, 70\%, 90\% for the distorted data at the percentages of 50\%, 30\%, 10\% over the entire transmitted dataset. This suggests that the distorted simplices can be partially recovered by inference from adjacent simplices using simplicial convolution from the topology. Furthermore, the distortion can affect the other correctly received simplices when rebuilding from embeddings. This explains the results that $p\%$ simplices remain distorted. The accuracy at SNR = 10 dB and 5dB is at around 10\% lower than SNR = 20 dB, with slow increase after more communication and training rounds. The fluctuation is also reduced after 200 rounds. These observations suggest that the SAE approach with high order semantic embedding effectively reduces the distortion and noise impact during transmission and infers the damaged data.

\section{Conclusions and Future Works} \label{sec:conclusion}
In this paper, we investigated semantic communications over a topological space of simplicial complexes. We proposed a simplicial autoencoder convolutional neural network to encode both simplicial feature and topology into a latent vector. A bilinear decoder was used to decode the simplicial Laplacians from the received embeddings, followed by a simplicial convolutional decoder to extract the features. The simplicial autoencoder was trained during communication of simplicial complex data between the transceivers. 

Simulation results from a coauthorship dataset showed that the proposed method effectively rebuilds simplicial feature and infers missing simplices with a mean accuracy up to 95\%. In contrast, the conventional linear autoencoder operating on raw data failed, unless being trained on the entire dataset. Furthermore, we demonstrated that the proposed the method can infer distorted simplices at the receiver, and achieves the accuracy level of a single agent simplicial convolution, without prior structural knowledge. The high order semantic embedding provides stable performance under different levels of data distortion and channel noise. The results suggested that the simplicial convolutional autoencoder enables the transceiver to infer data on topological space, making communication more effective and reliable than transmitting the observed data. 

Learning over a topological space has the potential to make semantic communications more theoretically grounded, by means of learning a common concept and topological space to represent observations. An important direction is to study effective algebraic operations on the complex topology, such as (un)convolutions on images to (up)downsampling the complex to a low dimensional representation. Furthermore, generalizing the semantic topological learning to multi-modality information (i.e. language, image, point-cloud), and reasoning on multi-tasks is an important issue left for future work. 
% training distributed semantic models via multi-agent communication, with an adaptive embedding optimized for dynamic semantic and physical channel distortion is an important issue left for future work. 

\bibliographystyle{IEEEtran}
\bibliography{ref}

% Generated by IEEEtran.bst, version: 1.14 (2015/08/26)
\begin{thebibliography}{10}
\providecommand{\url}[1]{#1}
\csname url@samestyle\endcsname
\providecommand{\newblock}{\relax}
\providecommand{\bibinfo}[2]{#2}
\providecommand{\BIBentrySTDinterwordspacing}{\spaceskip=0pt\relax}
\providecommand{\BIBentryALTinterwordstretchfactor}{4}
\providecommand{\BIBentryALTinterwordspacing}{\spaceskip=\fontdimen2\font plus
\BIBentryALTinterwordstretchfactor\fontdimen3\font minus
  \fontdimen4\font\relax}
\providecommand{\BIBforeignlanguage}[2]{{%
\expandafter\ifx\csname l@#1\endcsname\relax
\typeout{** WARNING: IEEEtran.bst: No hyphenation pattern has been}%
\typeout{** loaded for the language `#1'. Using the pattern for}%
\typeout{** the default language instead.}%
\else
\language=\csname l@#1\endcsname
\fi
#2}}
\providecommand{\BIBdecl}{\relax}
\BIBdecl

\bibitem{Seo2021}
\BIBentryALTinterwordspacing
H.~Seo, J.~Park, M.~Bennis, and M.~Debbah, ``Semantics-native communication
  with contextual reasoning,'' 2021. [Online]. Available:
  \url{https://arxiv.org/abs/2108.05681}
\BIBentrySTDinterwordspacing

\bibitem{Xie2021}
H.~Xie, Z.~Qin, G.~Y. Li, and B.-H. Juang, ``Deep learning enabled semantic
  communication systems,'' \emph{IEEE Trans. Signal Process.}, vol.~69, pp.
  2663--2675, 2021.

\bibitem{Zhou2021}
\BIBentryALTinterwordspacing
Q.~Zhou, R.~Li, Z.~Zhao, C.~Peng, and H.~Zhang, ``Semantic communication with
  adaptive universal transformer,'' 2021. [Online]. Available:
  \url{https://arxiv.org/abs/2108.09119}
\BIBentrySTDinterwordspacing

\bibitem{Lu2022}
K.~Lu, Q.~Zhou, R.~Li, Z.~Zhao, X.~Chen, J.~Wu, and H.~Zhang, ``Rethinking
  modern communication from semantic coding to semantic communication,''
  \emph{IEEE Wireless Commun.}, pp. 1--13, 2022.

\bibitem{Jiang_2022}
P.~Jiang, C.-K. Wen, S.~Jin, and G.~Y. Li, ``Deep source-channel coding for
  sentence semantic transmission with {HARQ},'' \emph{{IEEE} Trans. Commun.},
  pp. 1--1, 2022.

\bibitem{Dai2021}
\BIBentryALTinterwordspacing
J.~Dai, S.~Wang, K.~Tan, Z.~Si, X.~Qin, K.~Niu, and P.~Zhang, ``Nonlinear
  transform source-channel coding for semantic communications,'' 2021.
  [Online]. Available: \url{https://arxiv.org/abs/2112.10961}
\BIBentrySTDinterwordspacing

\bibitem{Weng2022}
\BIBentryALTinterwordspacing
Z.~Weng, Z.~Qin, X.~Tao, C.~Pan, G.~Liu, and G.~Y. Li, ``Deep learning enabled
  semantic communications with speech recognition and synthesis,'' 2022.
  [Online]. Available: \url{https://arxiv.org/abs/2205.04603}
\BIBentrySTDinterwordspacing

\bibitem{Bronstein_2017}
M.~M. Bronstein, J.~Bruna, Y.~LeCun, A.~Szlam, and P.~Vandergheynst,
  ``Geometric deep learning: Going beyond euclidean data,'' \emph{{IEEE} Signal
  Process. Mag.}, vol.~34, no.~4, pp. 18--42, July 2017.

\bibitem{Feng2018}
\BIBentryALTinterwordspacing
Y.~Feng, H.~You, Z.~Zhang, R.~Ji, and Y.~Gao, ``Hypergraph neural networks,''
  2018. [Online]. Available: \url{https://arxiv.org/abs/1809.09401}
\BIBentrySTDinterwordspacing

\bibitem{Ebli2020}
S.~Ebli, M.~Defferrard, and G.~Spreemann, ``Simplicial neural networks,''
  \emph{ArXiv}, vol. abs/2010.03633, 2020.

\bibitem{Hajij2020}
\BIBentryALTinterwordspacing
M.~Hajij, K.~Istvan, and G.~Zamzmi, ``Cell complex neural networks,'' 2020.
  [Online]. Available: \url{https://arxiv.org/abs/2010.00743}
\BIBentrySTDinterwordspacing

\bibitem{Belfiore2021}
\BIBentryALTinterwordspacing
J.-C. Belfiore and D.~Bennequin, ``Topos and stacks of deep neural networks,''
  2021. [Online]. Available: \url{https://arxiv.org/abs/2106.14587}
\BIBentrySTDinterwordspacing

\bibitem{HORAK2013303}
D.~Horak and J.~Jost, ``Spectra of combinatorial laplace operators on
  simplicial complexes,'' \emph{Advances in Mathematics}, vol. 244, pp.
  303--336, 2013.

\bibitem{Hacker2020}
\BIBentryALTinterwordspacing
C.~Hacker, ``k-simplex2vec: a simplicial extension of node2vec,'' 2020.
  [Online]. Available: \url{https://arxiv.org/abs/2010.05636}
\BIBentrySTDinterwordspacing

\bibitem{Lerique2019}
\BIBentryALTinterwordspacing
S.~Lerique, J.~L. Abitbol, and M.~Karsai, ``Joint embedding of structure and
  features via graph convolutional networks,'' 2019. [Online]. Available:
  \url{https://arxiv.org/abs/1905.08636}
\BIBentrySTDinterwordspacing

\bibitem{Ammar2018}
\BIBentryALTinterwordspacing
W.~Ammar, D.~Groeneveld, C.~Bhagavatula \emph{et~al.}, ``Construction of the
  literature graph in semantic scholar,'' 2018. [Online]. Available:
  \url{https://arxiv.org/abs/1805.02262}
\BIBentrySTDinterwordspacing

\end{thebibliography}

\end{document}